\documentclass[fleqn,twoside]{article}
\usepackage{espcrc2}

\usepackage{graphicx}
\usepackage[figuresright]{rotating}
\usepackage{amsmath}
\def\gsim{\raise0.3ex\hbox{$>$\kern-0.75em\raise-1.1ex\hbox{$\sim$}}}
\def\lsim{\raise0.3ex\hbox{$<$\kern-0.75em\raise-1.1ex\hbox{$\sim$}}}

\title{Astrophysical consequences of the OPERA superluminal neutrino}

\author{Luis Gonzalez-Mestres\address{LAPP, Universit\'e de Savoie, CNRS/IN2P3, B.P. 110, 74941 Annecy-le-Vieux Cedex, France}}

\begin{document}

\begin{abstract}
A simple discussion of the recent OPERA result on the apparent critical speed of the muon neutrino is presented. We point out in particular some of the possible consistency problems of such an interpretation of the OPERA data with respect to well-established astrophysical observations.
\vspace{1pc}
\end{abstract}

\maketitle

\section{Introduction}

After MINOS, \cite{MINOS}, OPERA \cite{OPERA} has produced data suggesting that the critical speed of muon neutrinos in vacuum can be larger than that of light. The values of the measured relative differences are :
\vskip 2mm
- (5.1 $\pm $ 2.9) x 10$^{-5}$ (MINOS, 2007)
\vskip 2mm
- (2.48 $\pm $ 0.28) x 10$^{-5}$ (OPERA, 2011)
\vskip 2mm
These figures are much stronger than the prediction of any standard extrapolation from Lorentz symmetry violation (LSV) generated at the Planck scale \cite{Gonzalez-Mestres2010a}. Therefore, a first question requiring a close study is that of the consistency of these data with existing and well-established results from particle physics experiments and astrophysical observations. 

In particular, astrophysical bounds on possible critical speed anomalies with respect to the speed of light appear to be much more stringent than suggested by the recent OPERA data. It therefore seems necessary to study more closely the possible consistency of the OPERA result. 

\section{Effects on cosmic neutrinos}

In the high-energy approximation, using a kinematics of the Lorentz type, the energy $E$ of a particle X with critical speed in vacuum $c_X$, mass $m_X$ and momentum $p$ can be written as :
\begin{equation}
E ~ \simeq ~ p~c_X~+~m_X ^2~c_X ^3~(2~p)^{-1}
\end{equation}
Taking $c$ to be the speed of light, and assuming that charged leptons have the same critical speed in vacuum $c$, the emission of a neutrino with critical speed in vacuum $c_{\nu }$ = $c$ + $\delta c~ (\nu )$ and ~$\delta c$~ $(\nu )~>~0$ would cost extra energy as compared to the situation where $c_{\nu }$ = $c$. 

In the case of pion decay and with the above hypotheses, this extra energy $\delta E$ = $p_{\nu }$ $\delta c$ $(\nu )$, where $p_{\nu }$ is the neutrino momentum, can only be provided by the pion mass term $m_{\pi} ^2~c_{\pi} ^3~(2~p_{\pi})^{-1}$ ($m_{\pi}$ = charged pion mass, $p_{\pi}$ = pion momentum, $c_{\pi}$ = charged pion critical speed in vacuum). 

Therefore, if one assumes $c_{\pi}$ = $c$, the emission of such a neutrino will in any case be precluded by kinematics if $p_{\nu }$ $\delta c$ ($\nu $) $>~ m_{\pi} ^2~c ^3~(2~p_{\nu })^{-1}$. This leads to the equation :
\begin{equation}
p_{\nu } ^2~\leq ~m_{\pi} ^2~c ^3~[2~\delta c (\nu )]^{-1}
\end{equation}
Taking the OPERA value $\delta c (\nu )$ $c^{-1}$ = 2.5 x $10^{-5}$, one gets :
\begin{equation}
p_{\nu } ~\leq ~ 20~GeV/c
\end{equation}
Taking into account the mass term of the produced muon, the bound actually becomes $p_{\nu } ~\leq ~ 14~GeV/c$, so that decays of charged pions with a critical speed $c_{\pi}$ = $c$ cannot produce muon neutrinos with energies larger than $\simeq 14$ GeV. 

Such an effect cannot be compensated by a critical speed lower than $c$ for charged leptons, as in this case a high-energy photon would spontaneously decay into charged lepton pairs. Taking $\delta c (\mu )$ = - $\delta c$ $(\nu )$, where $\delta c (\mu )$ = $c_{\mu}~-~c$ and $c_{\mu}$ is the muon critical speed in vacuum, would lead to photon spontaneous decay into a $\mu ^+ ~\mu ^-$ pair at energies above $\simeq ~40 ~GeV$.

As OPERA reports a similar result for neutrinos with an average energy of 42.9 GeV, the only way out seems to be a positive value for $\delta c$ $(\pi )$ = $c_{\pi}$ - $c$ not much smaller than $\delta c$ $(\nu )$. However, this would allow for spontaneous photon emission by charged pions and imply a propagation of the critical speed anomaly to the hadronic sector, including the proton.

If $\delta c$ $(\pi )$ = $\delta c$ $(\nu )$, spontaneous photon emission by a charged pion would be allowed for pion energies above 20 GeV. Taking $\delta c$ $(\pi )$ = 10$^{-5} ~c$ would lead to spontaneous photon emission at pion energies above $\simeq $ 32 GeV. Because of its electromagnetic coupling, we expect this channel to be the leading one for pion decay at very high energy astrophysical sources. 

In all cases, on the grounds of the recent OPERA results, a strong suppression of very high-energy neutrinos produced through pion decay should be expected.

Another related issue is that of electron-positron pair emission. As the critical speed $c_e$ of these particles must remain close to that of light in order to avoid their own spontaneous decays, the OPERA data would imply the possibility of $e^+ ~e^-$ pair emission by neutrinos at energies $p_\nu ~c$ ($m_{e}$ = electron mass):
\begin{equation}
p_\nu ~c ~\geq ~ m_{e} ~c ^{3/2}~\delta c (\nu )^{-1/2} ~\simeq ~110~MeV
\end{equation}
A decay channel that, in any case, cannot be ignored at astrophysical energies and distances when considering cosmic neutrino fluxes. 

A similar calculation for 100 TeV neutrinos with an arbitrary value of $\delta c (\nu )$ leads to an allowed $e^+ ~e^-$ pair emission for $\delta c (\nu )$ $c^{-1}~ \gsim ~10^{-16}$. 

Therefore, even if a violation of Lorentz symmetry can explain the lack of a 100 TeV neutrino signal from gamma-ray bursts, the relevant figures have little to do with the strong LSV inferred from the OPERA signal that would, on the contrary, suppress pion-emitted neutrino fluxes at much lower energies.

\section{Implications for cosmic hadrons}

Similarly, it seems very difficult to reconcile the OPERA result with hadronic cosmic-ray physics. 

As a significant critical speed anomaly for pions, $\delta c$ $(\pi )$ of the order $\approx 10^{-5}~c$, seems in any case necessary to explain the data, such an anomaly will in principle be transmitted to all hadrons through standard couplings. 

We therefore expect the proton critical speed anomaly to be at least of the order $\approx 10^{-6}~c$. But such a result leads to obvious phenomenological problems :

- If $\delta c$ $(p)$ = $c_p$ - $c$ ($c_p$ = proton critical speed) is positive and $\approx 10^{-6}~c$, the same kind of calculations as before lead to a proton spontaneous decay by photon emission allowed above an energy $E ~\simeq  ~ 700 ~ GeV$, eight orders of magnitude below the highest observed cosmic-ray energies. 

- Similarly, with a negative value of $\delta c$ $(p)$ $\approx $ - $10^{-6}~c$, a photon with an energy as low as 1 TeV would be able to decay into a proton-antiproton pair.

- Taking $\delta c$ $(p)$ of the order $\approx 10^{-8}~c$ leads to thresholds higher by a factor of 10 for both processes. Such thresholds would still be clearly in conflict with observations.

As in the previous section, these figures would tend to suggest that the value of $\delta c$ $(\nu )$ obtained by OPERA is too far from possible realistic values to be considered as significant, and that an experimental problem may be at the origin of this result.

\section{Conclusion}

Crucial to this discussion was the fact that the OPERA result requires a positive critical speed anomaly for charged pions not too far from that of the muon neutrino. Otherwise, $40 ~GeV$ neutrinos could not have been produced by the CERN neutrino beam to Gran Sasso if the OPERA result is to be taken literally.

In all cases, neutrino production by pion decay at high energy appears to require for the pion the same kind of critical speed anomaly as that of the emitted neutrino. Then, such an anomaly naturally propagates to hadrons. 

This situation has important unwanted astrophysical consequences and leads also to consistency problems from the point of view of Particle Physics.

It therefore seems that the OPERA data, interpreted as a result on the critical speed in vacuum of the muon neutrino, cannot be reconciled with astrophysical observations or even with standard accelerator particle physics. Further work on the subject is obviously required. 

In spite of the consistency problems described here, it would be worth exploring from a theoretical point of view situations with much weaker critical speed anomalies that would not correspond to the usual extrapolations from LSV at the Planck scale.

\end{document}